\documentclass[%
 amsmath,amssymb,
 aps, physrev, twocolumn
]{revtex4-2}

\usepackage{graphicx}
\usepackage{dcolumn}
\usepackage{bm}
\usepackage[normalem]{ulem}

\usepackage{physics}
\usepackage{marginnote}
\usepackage{xcolor}
\usepackage[colorlinks=true, linkcolor=black, citecolor=blue, urlcolor=blue]{hyperref}
\usepackage[nameinlink,noabbrev]{cleveref}

\usepackage{booktabs}
\usepackage{threeparttable}

\usepackage[textwidth=5cm]{todonotes}
\presetkeys{todonotes}{backgroundcolor=cyan!15, linecolor=cyan!50!black, textcolor=black}{}
\begin{document}

\preprint{APS/123-QED}

\title{\textbf{Perturbative Recovery of Stabilizer-Subspace Correlations for Quantum Computational Chemistry} 
}%
\author{Qianjun Yao}
\affiliation{
 School of Electronic Science and Engineering, Southeast University, Nanjing, 211189, China
}%

\author{Jie Liu}
\email{Email: liujie86@ustc.edu.cn}
\affiliation{
 Hefei National Laboratory, Hefei, 230088, China
}%

\author{He Li}
\email{Email: helix@seu.edu.cn}
\affiliation{%
  School of Electronic Science and Engineering, Southeast University, Nanjing, 211189, China
}%

\date{November 16, 2026}

\begin{abstract}
Stabilizer-based subspace methods can substantially reduce the quantum resources required for electronic structure calculations, but the resulting projection neglects Hamiltonian couplings to configurations outside the retained subspace. In this work, we introduce Stabilizer-Subspace Perturbation Theory (SSPT) to recover the associated correlation contributions, using the Contextual Subspace Variational Quantum Eigensolver (CS-VQE) as a representative application. The central idea is to exploit the Pauli terms eliminated by stabilizer projection as perturbative operators that generate an external manifold from the optimized reference state. Their energetic contributions are evaluated using an Epstein–Nesbet-type second-order correction or a resolvent-based formulation that retains couplings within the external space. By exploiting the factorized structure of the stabilizer-constrained reference state, the required matrix elements can be evaluated through Pauli expectation values on the reduced contextual register without explicitly reconstructing the full-space wave function. Molecular benchmarks demonstrate substantial reductions in projection-induced energy errors, allowing chemical precision to be achieved with fewer contextual qubits in several cases. These results establish stabilizer-guided perturbation theory as a promising strategy for improving the accuracy of resource-efficient quantum chemistry calculations.
\end{abstract}

\enlargethispage{1\baselineskip}
\maketitle

\section{Introduction}
Accurately describing electron correlation remains a central challenge in electronic structure theory~\cite{szabo1996modern,helgaker2000molecular}. Classical electronic structure methods, including selected configuration interaction (SCI)~\cite{Holmes2016,sharma2017semistochastic}, density matrix renormalization group (DMRG)~\cite{White1999,Chan2011}, and quantum Monte Carlo (QMC)~\cite{Booth2009}, have achieved remarkable success in describing strongly correlated molecular systems. Nevertheless, their computational costs can become substantial as the number of correlated orbitals increases, motivating the development of alternative approaches to electronic structure calculations. Quantum computing offers a potential alternative for addressing these challenges~\cite{AspuruGuzik2005,Kassal2008,Cao2019quantum}. Among the proposed quantum algorithms, the Variational Quantum Eigensolver (VQE)~\cite{peruzzo2014variational,mcclean2016theory,cerezo2021variational,tilly2022variational} has attracted considerable attention as a hybrid quantum-classical approach for predicting electronic structure properties. However, its practical applicability remains limited by available quantum resources, as the number of qubits required generally increases with the size of the active space, accompanied by growing circuit and measurement costs~\cite{wecker2015progress,mcardle2020quantum,gonthier2022measurements}.

To overcome these limitations, considerable efforts have been devoted to reducing the quantum resources required for electronic structure calculations. Representative approaches include symmetry-based qubit tapering~\cite{bravyi2017tapering,setia2020reducing}, effective Hamiltonian downfolding~\cite{metcalf2020resource,huang2023leveraging,zeng2025accurate}, entanglement forging~\cite{eddins2022doubling}, and contextual subspace methods~\cite{kirby2019contextuality,kirby2021contextual,weaving2023stabilizer}. In particular, the contextual subspace variational quantum eigensolver (CS-VQE) separates the Hamiltonian into a classically tractable noncontextual part and a contextual correction, enabling the variational calculation to be performed within a reduced Hilbert space defined by stabilizer constraints.

While the CS-VQE approach can capture essential multiconfigurational features associated with static correlation, accurate molecular energetics generally require a balanced treatment of both static and dynamic electron correlation. This is particularly important for quantities such as reaction barriers and relative electronic energies, where missing correlation contributions can significantly affect the predicted energy differences. In conventional electronic structure theory, multireference perturbation methods, including complete active space second-order perturbation theory (CASPT2)~\cite{andersson1992secondorder} and N-electron valence state second-order perturbation theory (NEVPT2)~\cite{angeli2001introduction,angeli2002nelectron}, provide efficient approaches to recovering dynamic correlation beyond active-space approximation. Similar strategies have been extended to hybrid quantum-classical calculations by combining quantum-computed reference states with classical perturbative corrections~\cite{ryabinkin2021aposteriori,tammaro2023nelectron,krompiec2022strongly,zhao2024enhancing,shirai2025enhancing,zheng2026perturbative}.

Unlike conventional active-space approximations, stabilizer-based subspace reduction introduces an additional source of correlation loss within a fixed orbital space. In CS-VQE, the variational wave function is constrained to a reduced Hilbert space defined by stabilizers derived from the noncontextual Hamiltonian. These constraints are not necessarily exact symmetries of the full Hamiltonian. Consequently, Pauli terms that anticommute with the enforced stabilizers vanish under projection, although they may still couple the optimized contextual reference state to configurations outside the retained subspace~\cite{weaving2023stabilizer,ralli2023unitary}. Such couplings are neglected by the reduced-space variational optimization and provide a natural starting point for perturbative corrections. These corrections therefore address correlation effects neglected by stabilizer projection, rather than those arising solely from excitations outside the chosen orbital space.

In this work, we introduce a Stabilizer-Subspace Perturbation Theory (SSPT) framework to recover correlation effects neglected by stabilizer-based subspace projections, with CS-VQE serving as a representative application. The central idea is to construct an external perturbative space by applying the Pauli operators removed during stabilizer projection to the optimized CS-VQE reference state. The resulting energy corrections are evaluated using Epstein--Nesbet-type second-order perturbation theory~\cite{epstein1926stark,sharma2017semistochastic} or a resolvent-based formulation~\cite{lowdin1962studies,burton2024rationale} that retains couplings within the selected perturbative space. By exploiting the factorized structure of the stabilizer-constrained reference state, the required matrix elements can be evaluated through Pauli expectation values on the reduced contextual register without explicitly constructing the full-space wave function. We assess the accuracy and robustness of SSPT through molecular benchmarks, including potential energy curves, size-consistency tests, and reaction barrier calculations. The results demonstrate that perturbative recovery of projection-induced correlation effects can substantially improve molecular energies while retaining the qubit savings provided by CS-VQE.

\section{Method}

\subsection{A brief introduction of CS-VQE}
A molecular Hamiltonian expressed in the Pauli basis is generally written as
\begin{equation}
\label{eq:h}
\hat{H} = \sum_i h_i \hat{P}_i,
\end{equation}
where $\hat{P}_i$ are Pauli strings and $h_i$ are real coefficients obtained from the one- and two-electron integrals through a fermion-to-qubit transformation~\cite{jordan1928paulische,bravyi2002fermionic,tranter2018comparison}. In CS-VQE~\cite{kirby2021contextual,weaving2023stabilizer}, the Hamiltonian is partitioned into a noncontextual part $\hat{H}_{\mathrm{NC}}$ and a remaining contextual part $\hat{H}_{\mathrm{C}}$,
\begin{equation}
\label{eq:h_partition}
\hat{H} = \hat{H}_{\mathrm{NC}} + \hat{H}_{\mathrm{C}}.
\end{equation}
The noncontextual Hamiltonian is constructed from a selected subset of Pauli terms whose commutation structure allows its energy to be determined via classical optimization. The optimized noncontextual solution determines a set of mutually commuting stabilizer constraints $\{\hat{G}_k\}$ together with fixed eigenvalues $g_k\in\{\pm1\}$. These constraints define a projector onto the selected stabilizer subspace,
\begin{equation}
\hat{\Pi}_{\mathbf{g}}=\prod_k \frac{\hat{I}+g_k \hat{G}_k}{2},
\end{equation}
and thereby partitions the full problem into a classically tractable noncontextual part and a residual contextual part requiring quantum computation. 

To construct the contextual-subspace Hamiltonian, a unitary transformation $\hat{U}$ is first introduced to map the stabilizer generators onto single-qubit Pauli operators acting on distinct qubits~\cite{kirby2021contextual,weaving2023stabilizer},
\begin{equation}
\label{eq:stabilizer_rotation}
\hat{U}\hat{G}_k\hat{U}^{\dagger}
=
\hat{Z}_{f(k)},
\end{equation}
where $f(k)$ specifies the qubit associated with the $k$th stabilizer generator and $\hat{Z}$ is the Pauli-Z operator. The transformed Hamiltonian is defined as
\begin{equation}
\label{eq:h_rot}
\widetilde{H}
=
\hat{U}\hat{H}\hat{U}^{\dagger}.
\end{equation}
The contextual-subspace Hamiltonian is then obtained by projecting onto the prescribed stabilizer eigenstates and tracing out the corresponding fixed qubits,
\begin{equation}
\label{eq:h_cs}
\mathcal{H}_{\mathrm{CS}}
=
\operatorname{Tr}_{\mathrm{stab}}
\left[
\hat{U}\hat{\Pi}_{\mathbf{g}}
\hat{H}\hat{\Pi}_{\mathbf{g}}
\hat{U}^{\dagger}
\right],
\end{equation}
where the partial trace is taken over the qubits indexed by $f(k)$ in the rotated representation. Throughout this work, $\hat{H}$ and $\widetilde{H}$ act on the full register in the original and rotated representations, respectively, whereas $\mathcal{H}_{\mathrm{CS}}$ acts only on the reduced contextual register. We use lowercase $|\psi_{\mathrm{CS}}\rangle$ for the reduced-register wave function and uppercase $|\Psi_{\mathrm{CS}}\rangle$ (or $|\widetilde{\Psi}_{\mathrm{CS}}\rangle$) for its full-register embedding; a tilde denotes a unitary rotation, not a reduction in Hilbert-space dimension.
In this sense, CS-VQE provides a systematic route from the noncontextual approximation toward the full variational problem, with the subspace dimension controlled by the number of qubits retained as contextual degrees of freedom.

A particularly important consequence of the stabilizer projection is that Pauli terms anticommuting with any of the enforced stabilizer generators vanish within the projected subspace. Specifically, if a Pauli operator $\hat{P}$ anticommutes with at least one generator $\hat{G}_k$, i.e.,
$\{\hat{P},\hat{G}_k\}=0$, then
\begin{equation}
\hat{\Pi}_{\mathbf{g}} \hat{P}
\hat{\Pi}_{\mathbf{g}}=0.
\end{equation}
For any state $|\psi_{\mathbf{g}}\rangle$ satisfying the stabilizer constraints
$\hat{G}_k|\psi_{\mathbf{g}}\rangle
= g_k|\psi_{\mathbf{g}}\rangle$ for all $k$,
the expectation value of such a Pauli operator vanishes,
\begin{equation}
\langle\psi_{\mathbf{g}}|\hat{P}|
\psi_{\mathbf{g}}\rangle=0.
\end{equation}
Consequently, when all stabilizer generators are Pauli strings, only Pauli terms commuting with every generator can contribute directly to the effective CS-VQE Hamiltonian. More generally, in the rotated Pauli representation, only Pauli components acting as either identity or Pauli-Z on each stabilized qubit survive the projection.

However, the anticommuting terms are not necessarily physically negligible, as they can couple the selected stabilizer subspace to its orthogonal complement. Given the complementary projector $\hat{Q}_{\mathbf{g}}=\hat{I}-\hat{\Pi}_{\mathbf{g}}$, the full Hamiltonian can then be decomposed as
\begin{equation}
\label{eq:h_decomposition}
\begin{aligned}
\hat{H} ={}&
\underbrace{
\hat{\Pi}_{\mathbf{g}}\hat{H}\hat{\Pi}_{\mathbf{g}}
}_{\mathrm{contextual}}
+
\underbrace{
\hat{Q}_{\mathbf{g}}\hat{H}\hat{Q}_{\mathbf{g}}
}_{\mathrm{complementary}}
\\
&+
\underbrace{
\hat{\Pi}_{\mathbf{g}}\hat{H}\hat{Q}_{\mathbf{g}}
+
\hat{Q}_{\mathbf{g}}\hat{H}\hat{\Pi}_{\mathbf{g}}
}_{\mathrm{coupling}}.
\end{aligned}
\end{equation}
The first term represents the Hamiltonian restricted to the contextual subspace, while the second describes the complementary (excluded) subspace. The remaining two terms account for the couplings between these subspaces, which are neglected in the CS-VQE calculation. Although these couplings do not contribute directly to the variational energy within the projected subspace, they can give rise to additional correlation energy. This observation provides a natural motivation for introducing perturbative corrections to recover the contributions neglected by the stabilizer projection.

\subsection{Perturbation theory}\label{sec:perturbation_theory}
Perturbation theory provides a systematic framework for improving approximate solutions to quantum many-body problems. Among various perturbative approaches, Rayleigh--Schr\"odinger perturbation theory (RSPT)~\cite{shavitt2002international, Møller1934note} expands the energy and wave function in powers of a perturbation around a solvable reference Hamiltonian. The full Hamiltonian is expressed as
\begin{equation}
\label{eq:pt_hamiltonian}
\hat{H}=\hat{H}_0+\lambda\hat{V},
\end{equation}
where $\hat{H}_0$ is the zeroth-order Hamiltonian, $\hat{V}$ is the perturbation, and $\lambda$ is a formal expansion parameter.

For the CS-VQE framework, a natural partition is obtained from the stabilizer projection,
\begin{equation}
\label{eq:h0_v}
\begin{aligned}
\hat{H}_0
={}&
\hat{\Pi}_{\mathbf{g}}\hat{H}
\hat{\Pi}_{\mathbf{g}}
+
\hat{Q}_{\mathbf{g}}\hat{H}
\hat{Q}_{\mathbf{g}},\\
\hat{V}
={}&
\hat{\Pi}_{\mathbf{g}}\hat{H}
\hat{Q}_{\mathbf{g}}
+
\hat{Q}_{\mathbf{g}}\hat{H}
\hat{\Pi}_{\mathbf{g}}.
\end{aligned}
\end{equation}
Here, the zeroth-order Hamiltonian preserves
the contextual and complementary subspaces,
while the perturbation describes the
couplings between them.

Taking the optimized CS-VQE state
$|\Psi_{\mathrm{CS}}\rangle$ as the reference,
with
$\hat{\Pi}_{\mathbf{g}}|\Psi_{\mathrm{CS}}\rangle
=|\Psi_{\mathrm{CS}}\rangle$,
the first-order energy correction vanishes,
\begin{equation}
\label{eq:e1}
E^{(1)}
=
\langle\Psi_{\mathrm{CS}}|
\hat{V}
|\Psi_{\mathrm{CS}}\rangle=0.
\end{equation}
Here, $|\Psi_{\mathrm{CS}}\rangle$ denotes the optimized CS-VQE reference state embedded in the original Hilbert space, whose explicit construction is given in Sec.~\ref{sec:C}. The leading nonvanishing contribution
therefore arises at second order.
Within RSPT, the second-order energy
correction is given by
\begin{equation}
\label{eq:e2}
\begin{aligned}
E^{(2)}
={}&
\langle\Psi_{\mathrm{CS}}|
\hat{\Pi}_{\mathbf{g}}\hat{H}
\hat{Q}_{\mathbf{g}}
\\
&\times
\left(
E_{\mathrm{CS}}
-\hat{Q}_{\mathbf{g}}\hat{H}
\hat{Q}_{\mathbf{g}}
\right)^{-1}
\\
&\times
\hat{Q}_{\mathbf{g}}\hat{H}
\hat{\Pi}_{\mathbf{g}}
|\Psi_{\mathrm{CS}}\rangle,
\end{aligned}
\end{equation}
where $E_{\mathrm{CS}}$ is the reference
energy obtained from CS-VQE, and the
inverse is evaluated within the excluded
subspace. In the practical implementation, one can consider an orthonormal basis $\{|\phi_\mu\rangle\}$ spanning the complementary subspace and define
\begin{equation}
\label{eq:full_external_matrices}
\begin{aligned}
(H_Q)_{\mu\nu}
&=\langle\phi_\mu|\hat{H}|\phi_\nu\rangle,\\
(v_Q)_\mu
&=\langle\phi_\mu|\hat{H}|\Psi_{\mathrm{CS}}\rangle.
\end{aligned}
\end{equation}
Equation~(\ref{eq:e2}) can then be written as
\begin{equation}
\label{eq:full_resolvent_matrix}
E_{\mathrm{res}}^{(2)}
=\mathbf{v}_Q^\dagger
\left(E_{\mathrm{CS}}I-H_Q\right)^{-1}
\mathbf{v}_Q.
\end{equation}
A less expensive alternative is motivated by second-order Epstein--Nesbet perturbation theory (ENPT2)~\cite{epstein1926stark,sharma2017semistochastic}, which neglects the off-diagonal elements of $H_Q$ and approximates the complementary-space Hamiltonian as
\begin{equation}
\label{eq:q_diagonal}
\hat{Q}_{\mathbf{g}}\hat{H}
\hat{Q}_{\mathbf{g}}
\approx
\sum_{\mu}
E_{\mu}
|\phi_{\mu}\rangle
\langle\phi_{\mu}|,
\end{equation}
where
$E_{\mu}=\langle\phi_{\mu}|\hat{H}
|\phi_{\mu}\rangle$.
The second-order correction then reduces to
\begin{equation}
\label{eq:cs_enpt2}
E_{\mathrm{CS}}^{(2)}
=
\sum_{\mu}
\frac{
|\langle\phi_{\mu}|\hat{H}|
\Psi_{\mathrm{CS}}\rangle|^2
}{
E_{\mathrm{CS}}-E_{\mu}
}.
\end{equation}
This expression accounts for the coupling between the optimized CS-VQE reference state and configurations in the complementary subspace, providing an Epstein--Nesbet-type perturbative correction to the energy obtained from the stabilizer-subspace Hamiltonian. Unlike the resolvent-based expression of Eq.~\eqref{eq:full_resolvent_matrix}, it neglects couplings between distinct external basis states and therefore depends on the chosen basis. In Sec.~\ref{sec:C}, we implement both treatments in a selected external manifold generated by the discarded Pauli operators, replacing the full complementary space by a tractable perturbative space.

\subsection{Stabilizer-Subspace Perturbative Correction}\label{sec:C}
After the CS-VQE optimization, the variational
state $|\psi_{\mathrm{CS}}\rangle$ is obtained
within the reduced contextual subspace.
The corresponding
full-register reference state in the transformed
representation can be written as
\begin{equation}
\label{eq:psi_rot}
|\widetilde{\Psi}_{\mathrm{CS}}\rangle
=
|\psi_{\mathrm{CS}}\rangle
\otimes
\bigotimes_{k\in\mathrm{stab}}|\psi_k\rangle,
\end{equation}
where $|\psi_k\rangle$ is the fixed eigenstate
of the corresponding single-qubit stabilizer
operator. The reference state in the original
Hilbert space is recovered through
\begin{equation}
\label{eq:psi_full}
|\Psi_{\mathrm{CS}}\rangle
=
\hat{U}^{\dagger}
|\widetilde{\Psi}_{\mathrm{CS}}\rangle.
\end{equation}
The reference energy is
preserved under the unitary transformation,
\begin{equation}
\label{eq:energy_consistency}
\begin{aligned}
E_{\mathrm{CS}}
&=
\langle\psi_{\mathrm{CS}}|
\mathcal{H}_{\mathrm{CS}}
|\psi_{\mathrm{CS}}\rangle \\
&=
\langle\widetilde{\Psi}_{\mathrm{CS}}|
\widetilde{H}
|\widetilde{\Psi}_{\mathrm{CS}}\rangle \\
&=
\langle\Psi_{\mathrm{CS}}|
\hat{H}
|\Psi_{\mathrm{CS}}\rangle.
\end{aligned}
\end{equation}

Although Eq.~(\ref{eq:psi_full}) formally
reconstructs the reference state in the full
Hilbert space, explicitly expanding the
corresponding statevector is not required.
Instead, the tensor-product structure in
Eq.~(\ref{eq:psi_rot}) allows the matrix
elements needed for perturbation theory
to be evaluated in the reduced contextual
representation.

For a Pauli string
$\hat{R}
=
\hat{R}_{\mathrm{sim}}
\otimes
\bigotimes_{k\in\mathrm{stab}}\hat{R}_k$,
its expectation value factorizes as
\begin{equation}
\label{eq:factorized_expectation}
\begin{aligned}
&
\langle\widetilde{\Psi}_{\mathrm{CS}}|
\hat{R}
|\widetilde{\Psi}_{\mathrm{CS}}\rangle
\\
&=
\langle\psi_{\mathrm{CS}}|
\hat{R}_{\mathrm{sim}}
|\psi_{\mathrm{CS}}\rangle
\prod_{k\in\mathrm{stab}}
\langle\psi_k|\hat{R}_k|\psi_k\rangle.
\end{aligned}
\end{equation}
This factorization provides the key computational advantage of the subspace implementation. Although the perturbative states span sectors outside the retained stabilizer subspace, their overlap, Hamiltonian, and coupling matrix elements can be expressed entirely in terms of Pauli expectation values evaluated on the contextual register, multiplied by analytically determined stabilizer contributions. The factorization is exact and introduces no additional approximation beyond the truncation of the perturbative manifold. Consequently, the required expectation values can be evaluated using the optimized contextual state and the known stabilizer eigenstates. 

To construct the perturbative space, 
Pauli terms $\{\widetilde{P}_i\}$ are selected from the
transformed Hamiltonian that anticommute
with at least one enforced stabilizer.
Applying these operators to the reconstructed
reference state generates a set of perturbative
states,
\begin{equation}
\label{eq:external_states}
|\widetilde{\Phi}_i\rangle
=
\widetilde{P}_i
|\widetilde{\Psi}_{\mathrm{CS}}\rangle.
\end{equation}
By construction, these states lie outside
the retained stabilizer subspace and are
orthogonal to the reference state.

The corresponding overlap matrix,
Hamiltonian matrix, and coupling vector
are defined as
\begin{equation}
\label{eq:pt_matrices}
\begin{aligned}
S_{ij}
&=
\langle\widetilde{\Phi}_i|
\widetilde{\Phi}_j\rangle,\\
(H_{\mathrm{pt}})_{ij}
&=
\langle\widetilde{\Phi}_i|
\widetilde{H}
|\widetilde{\Phi}_j\rangle,\\
b_i
&=
\langle\widetilde{\Phi}_i|
\widetilde{H}
|\widetilde{\Psi}_{\mathrm{CS}}\rangle.
\end{aligned}
\end{equation}
All these quantities can be expressed as
sums of Pauli expectation values and
evaluated using the factorization in
Eq.~(\ref{eq:factorized_expectation}),
without explicitly constructing the
full-space reference statevector.

The linearly dependent perturbative states are removed through canonical orthonormalization of the overlap matrix,
\begin{equation}
\label{eq:canonical_orth}
S=V\Lambda V^\dagger,
\qquad
X=V_r\Lambda_r^{-1/2},
\end{equation}
where only eigenvectors corresponding to eigenvalues above a chosen threshold are retained. The Hamiltonian matrix and coupling vector in the orthonormalized perturbative space are then given by
\begin{equation}
\label{eq:orth_matrices}
\overline{H}
=
X^\dagger H_{\mathrm{pt}}X,
\qquad
\overline{\mathbf{b}}
=
X^\dagger\mathbf{b}.
\end{equation}
Applying the resolvent-based treatment of Sec.~\ref{sec:perturbation_theory} to this orthonormalized, selected external space gives
\begin{equation}
\label{eq:subspace_resolvent}
E_{\mathrm{res}}^{(2)}
=
\overline{\mathbf{b}}^\dagger
\left(
E_{\mathrm{CS}}I-\overline{H}
\right)^{-1}
\overline{\mathbf{b}}.
\end{equation}
Further, within the Epstein--Nesbet-type diagonal approximation, the second-order correction is evaluated as
\begin{equation}
\label{eq:subspace_enpt2}
E_{\mathrm{EN}}^{(2)}
=
\sum_j
\frac{|\overline{b}_j|^2}
{E_{\mathrm{CS}}-\overline{H}_{jj}}.
\end{equation}

The key advantage of this formulation is that the perturbative corrections can be evaluated without explicitly reconstructing the full $2^N$-dimensional reference state. As shown above, all matrix elements required for the perturbative corrections can be expressed as Pauli expectation values over the optimized contextual state, together with analytically determined contributions from the fixed stabilizer qubits. Consequently, the quantum calculations involve only the reduced contextual register, avoiding the exponential memory overhead associated with explicit full-space statevector representations.

Furthermore, the perturbative space can be systematically enlarged or truncated simply by the squared coefficient of the discarded Pauli operators $h_{i}^2$ in the perturbative operator, or by selecting a subset of the discarded Pauli operators according to their coupling strengths to the CS-VQE reference.
For each candidate operator $\widetilde P_i$, we define the coupling matrix element $V_{0i}=\langle\widetilde\Psi_{\mathrm{CS}}|\widetilde V|\widetilde\Phi_i\rangle$, where $\widetilde V$ denotes the perturbation expressed in the rotated stabilizer basis. Since the candidate external states lie outside the retained stabilizer subspace, this coupling can equivalently be evaluated as $V_{0i}=\langle\widetilde\Psi_{\mathrm{CS}}| \tilde{H}|\widetilde\Phi_i\rangle$. The candidate operators are ranked in descending order of $h_{i}^2$ or $|V_{0i}|^2$, and the leading $M$ operators are retained before orthonormalization of the generated external states. Because these configurations can be nonorthogonal or linearly dependent, their cumulative coupling weight should not be interpreted as the fraction of the second-order energy correction recovered. For $M$ selected perturbative operators and $L$ Pauli terms in the transformed Hamiltonian, the construction of the overlap and Hamiltonian matrices requires at most $O(M^2L)$ Pauli expectation-value contributions. After removal of linearly dependent directions and orthonormalization, the resulting matrix elements can be used in either the Epstein--Nesbet-type or resolvent-based correction to recover correlation effects neglected by CS-VQE.

\subsection{Workflow of SSPT}

The overall workflow of SSPT consists of five main steps, as summarized below.

\begin{enumerate}

\item \textbf{CS-VQE optimization.} We first perform a CS-VQE calculation to obtain the optimized variational state $|\psi_{\mathrm{CS}}\rangle$ and the corresponding reference energy $E_{\mathrm{CS}}$. The reference state $|\widetilde{\Psi}_{\mathrm{CS}}\rangle$ is then constructed using the fixed stabilizer eigenstates and the optimized contextual wave function, as described in Sec.~II.C.

\item \textbf{Perturbative operator selection.} From the transformed Hamiltonian $\widetilde{H}$, we identify the Pauli terms that anticommute with at least one enforced stabilizer generator. These terms vanish under the contextual-subspace projection but can couple the reference state to the complementary subspace. The corresponding Pauli operators $\{\widetilde{P}_i\}$ are selected to construct the perturbative space. To control the computational cost, the operators can be ranked according to the magnitudes of their Hamiltonian squared coefficients $h_{i}^2$.

\item \textbf{Construction of the perturbative space.} The selected Pauli operators are applied to the CS-VQE reference state to generate a set of candidate external states,
\begin{equation}
\label{eq:procedure_external}
|\widetilde{\Phi}_i\rangle =
\widetilde{P}_i|\widetilde{\Psi}_{\mathrm{CS}}\rangle.
\end{equation}
These states span a truncated perturbative space associated with the couplings neglected by the stabilizer projection.

\item \textbf{Matrix evaluation and orthonormalization.} We construct the overlap matrix, Hamiltonian matrix, and coupling vector using the factorized Pauli expectation values derived in Sec.~II.C. Since the generated external states are generally nonorthogonal and may be linearly dependent, canonical orthonormalization is performed on the overlap matrix. Eigenvectors associated with eigenvalues below a numerical threshold $\epsilon$ are discarded, yielding a linearly independent orthonormal basis and the corresponding transformed Hamiltonian matrix $\overline{H}$ and coupling vector $\overline{\mathbf{b}}$.

\item \textbf{Perturbative energy correction.} Finally, the second-order energy correction is evaluated using either the Epstein--Nesbet-type approximation or the resolvent-based formulation derived in Sec.~II.C. The former employs diagonal energy denominators, whereas the latter retains the couplings among the selected external states. The corrected ground-state energy is obtained as
\begin{equation}
\label{eq:corrected_energy}
E_{\mathrm{SSPT}}
=
E_{\mathrm{CS}}+E^{(2)},
\end{equation}
where $E^{(2)}$ denotes the correction calculated using either formulation.

\end{enumerate}

The procedure provides a post-variational correction to the CS-VQE energy without increasing the number of qubits required for the variational optimization. The size of the perturbative space can be controlled independently through the selection of Pauli operators, allowing the computational cost and accuracy of the correction to be balanced.
Hereafter, we refer to the Epstein--Nesbet-type and resolvent-based formulations as SSPT-D and SSPT, respectively.

\section{Results}
\label{sec:results}

We evaluate the accuracy, numerical convergence, and practical applicability of SSPT-D and SSPT across a broader range of molecular systems. For the initial LiH and BeH$^+$ benchmarks, the molecular geometries and qubit Hamiltonians are adopted from the source code accompanying Ref.~\cite{weaving2023stabilizer}. These Hamiltonians were originally constructed using molecular data from the Computational Chemistry Comparison and Benchmark Database (CCCBDB)~\cite{johnson2020cccbdb} and processed with IBM's Qiskit Nature~\cite{matthew_treinish_2023_8190968}. 

The CS-VQE calculations are performed using the qubit-ADAPT-VQE~\cite{grimsley2019adaptive,tang2021qubitadaptvqe}, implementation in the Symmer package~\cite{ralli2022symmer}. The initial state is the Hartree--Fock reference after qubit tapering and contextual-subspace projection. The operator pool consists of individual Pauli strings derived from the qubit-mapped single and double excitation operators and subjected to the same tapering and projection. Each pool element is assigned a unit coefficient, with its rotation angle treated as an independent variational parameter. At each ADAPT iteration, the generator with the largest absolute energy derivative is appended to the ansatz. Its parameter is initialized to zero, and all parameters are jointly reoptimized using BFGS, with derivatives evaluated using the parameter-shift rule. The ADAPT procedure terminates when the largest absolute pool derivative falls below $10^{-3}$~Hartree, the energy change between successive ADAPT iterations falls below $10^{-10}$~Hartree, or the specified maximum number of iterations is reached. All VQE expectation values are evaluated using noiseless classical statevector simulation without finite-shot sampling.

To further examine the performance of SSPT across different molecular geometries, we calculate the potential energy curves of LiH and BeH$^+$. PySCF~\cite{sun2020recent} is employed to construct the molecular systems and obtain Hartree--Fock and full configuration interaction (FCI) reference energies, while OpenFermion~\cite{mcclean2020openfermion} is used to map the corresponding fermionic Hamiltonians onto qubit operators through the Jordan--Wigner transformation.

To assess the applicability of the factorized implementation, we extend the single-point benchmarks to a broader set of molecules and investigate the size-consistency behavior using two well-separated H$_2$ fragments.  
We further evaluate the performance of SSPT in predicting chemical reaction barriers by considering the decomposition of formaldehyde. For this application, the qubit Hamiltonians are constructed using CASSCF(8e,8o) orbitals in the cc-pVDZ basis, and the calculated reaction barriers are compared with the corresponding active-space CASSCF reference. These additional benchmarks examine the accuracy and robustness of SSPT for molecular energy calculations while retaining the reduced qubit requirements of CS-VQE.

\subsection{Accuracy}
\label{subsec:A}

\begin{figure}[!h]
    \centering
    \hspace{-5mm}
    \includegraphics[width=0.9\linewidth]{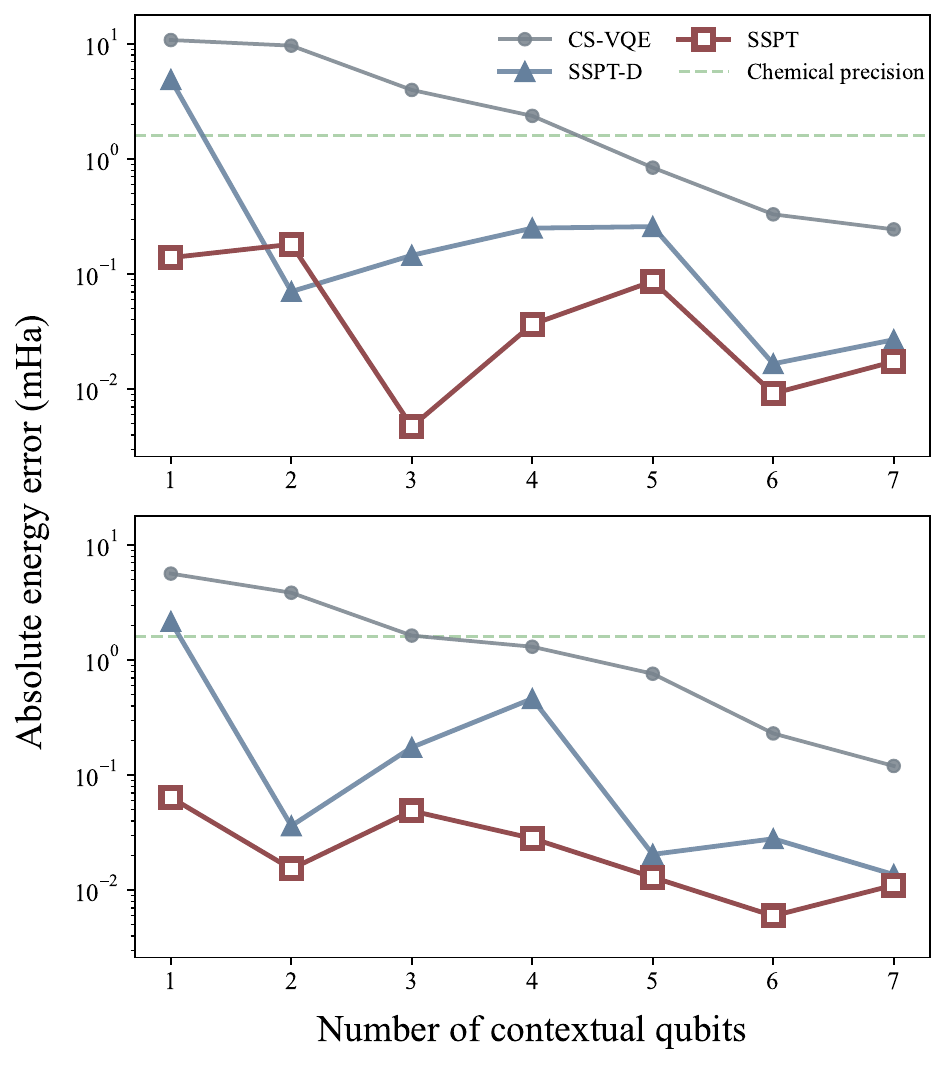}
    \caption{Absolute energy errors relative to the FCI reference are shown as a function of the number of qubits retained in the contextual subspace. The upper and down panels show the results of BeH$^+$ and LiH, respectively.}
    \label{fig:range_qubit}
\end{figure}

Figure~\ref{fig:range_qubit} compares the single-point energy errors obtained from bare CS-VQE, SSPT-D, and SSPT for LiH and BeH$^+$. Both perturbative treatments reduce the absolute energy errors at every reported contextual-subspace size. For BeH$^+$ with one contextual qubit, the CS-VQE error of $10.82$~milliHartree (mHa) is reduced to $4.90$~mHa by SSPT-D and $0.14$~mHa by SSPT. For LiH, the corresponding errors are $5.65$, $2.16$, and $0.06$~mHa, respectively. Taking $1.6$~mHa as the chemical-accuracy threshold, bare CS-VQE requires five contextual qubits for BeH$^+$ and four for LiH, whereas SSPT reaches this threshold with only one qubit for either molecule. SSPT-D requires two qubits for either molecule. Thus, SSPT reduces the minimum contextual-register size required for chemical precision by four qubits for BeH$^+$ and three for LiH. SSPT also gives smaller absolute errors than SSPT-D at every reported point except one at 2-qubit for BeH$^+$. For example, with three contextual qubits for BeH$^+$, the CS-VQE, SSPT-D, and SSPT errors are $3.98$, $0.15$, and $0.005$~mHa, respectively. 

It is worth mentioning that neither perturbative treatment guarantees monotonic improvement as the contextual register grows: the BeH$^+$ SSPT error increases from $0.005$~mHa at three qubits to $0.09$~mHa at five qubits, while remaining well below chemical precision. The signed errors further show that SSPT slightly overcorrects at one qubit, yielding energies below FCI by $0.14$~mHa for BeH$^+$ and $0.06$~mHa for LiH; SSPT-D remains above FCI at all reported points. At seven contextual qubits, the SSPT-D and SSPT errors are $0.03$ and $0.02$~mHa for BeH$^+$, and $0.014$ and $0.011$~mHa for LiH, compared with CS-VQE errors of $0.24$ and $0.12$~mHa, respectively. In the full-qubit limit, no contextual stabilizer constraints remain, so the corresponding external perturbative space disappears and both corrections vanish by construction. This limiting property does not imply monotonic convergence of the corrected energies or their absolute errors at intermediate register sizes.

\subsection{Convergence and resource requirement}
\label{subsec:B}

To examine convergence with respect to the selected perturbative operators, we fix the contextual register at two qubits and increase the number of retained Pauli terms in increments of ten, additionally including the points corresponding to the $98\%$ cumulative-coupling threshold and the complete candidate set. The operators are ranked in descending order of $h_{i}^2$ or $|V_{0i}|^2$, as defined above. Recomputing the SVD of the expanding matrix of Pauli-generated external-state vectors can rotate the orthonormal basis even when the underlying subspace remains unchanged, leading to fluctuations in the diagonal SSPT-D energy correction. To mitigate these basis-rotation-induced fluctuations, we freeze the existing basis independently for each term ordering once the normalized cumulative squared weight first reaches or exceeds 98\%. Subsequent excitation vectors $|\widetilde{\Phi}_i\rangle = \widetilde{P}_i|\widetilde{\Psi}_{\mathrm{CS}}\rangle$ are projected onto the orthogonal complement of the retained subspace. SVD is then applied to each batch of projected vectors, retaining only directions with singular values exceeding $10^{-10}$. The resulting orthonormal vectors are appended to the existing basis without modifying any previously retained vectors. To further examine how excitation-space truncation and basis rotations affect the numerical stability of SSPT-D, we present additional results for freezing thresholds of 50\%, 95\%, and 100\% in Appendix~\ref{appendix:basis_freezing_threshold}, which also provides a justification for our choice of the 98\% threshold. 

\begin{figure}[!h]
    \centering
    \hspace{-5mm}
    \includegraphics[width=1\linewidth]{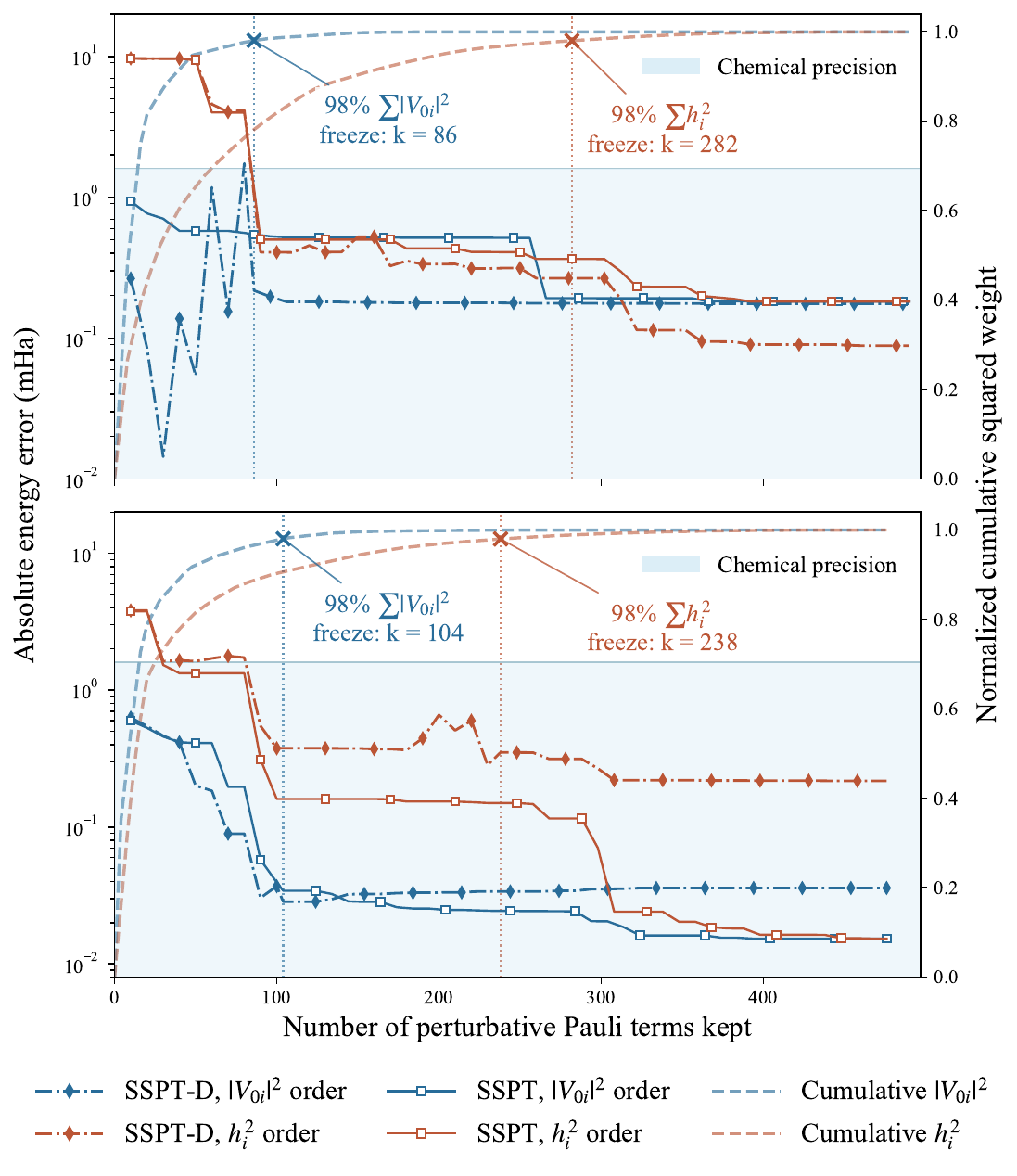}
    \caption{Absolute energy errors of SSPT-D and SSPT relative to the FCI reference as a function of the number of retained perturbative Pauli terms, using a two-qubit contextual-subspace reference state. The upper and lower panels correspond to $\mathrm{BeH}^{+}$ and $\mathrm{LiH}$, respectively. Terms are retained in descending order of $|V_{0i}|^2$ or $h_{i}^2$. Energy errors are evaluated at 10-term intervals within each SVD stage. The secondary y-axis shows the normalized cumulative sum of the corresponding squared weights. For each ordering, the cross ($\times$) marks the first term count at which the cumulative fraction reaches or exceeds 98\%. }
    \label{fig:convergence}
\end{figure}

Figure~\ref{fig:convergence} shows that both corrections reach chemical precision with relatively few operators. For the $|V_{0i}|^2$ ordering, with only ten terms, the absolute SSPT-D and SSPT errors are $0.26$ and $0.93$~mHa for BeH$^+$, and $0.63$ and $0.60$~mHa for LiH, respectively. However, their subsequent behavior differs substantially. For BeH$^+$, the SSPT-D error decreases to approximately $0.01$~mHa at 30 terms but increases to $0.18$~mHa when all 490 candidate terms are retained. For LiH, its error reaches approximately $0.028$~mHa at 104 terms before increasing to $0.036$~mHa for the complete set of 476 terms.
The non-monotonic SSPT-D behavior requires care in interpretation. The coupling-based selection criterion does not account for energy denominators. Moreover,  adding operators does not simply append fixed contributions to the sum in Eq.~\ref{eq:subspace_enpt2}, as the candidate external states are reorthonormalized at each truncation level before reaching the accumulative threshold, which can rotate the existing orthonormal basis and alter both the coupling components and the diagonal Hamiltonian elements. Because the diagonal approximation is generally not invariant under such basis rotations, its correction can change even when additional Pauli operators do not enlarge the external subspace. For example, in the BeH$^+$ calculation, increasing the retained set from 50 to 60 terms leaves the perturbative-space dimension at 21 and its span unchanged to numerical precision, yet changes the signed SSPT-D error from $-0.05$ to $-1.17$~mHa. The SSPT error remains $0.58$~mHa over the same interval. Thus, these fluctuations cannot be attributed solely to the inclusion of new near-degenerate external states. Furthermore, since second-order corrected energies are not variational upper bounds, even a decreasing energy need not produce a decreasing absolute error relative to FCI.

By retaining the full projected Hamiltonian within the selected external space, the resolvent expression in Eq.~\ref{eq:subspace_resolvent} is invariant under orthonormal basis rotations within that space. This removes the basis dependence associated with the diagonal approximation. In the present benchmarks, SSPT using the $h_{i}^2$ and $|V_{0i}|^2$ ordering exhibits an approximately monotonic decrease in absolute error, reaching $0.18$~mHa for BeH$^+$ and $0.02$~mHa for LiH with the complete candidate sets. Plateau regions can arise when added operators generate linearly dependent configurations and therefore leave the external span unchanged, or when newly accessible directions have little effect on the correction. These observations support the numerical stability of the resolvent treatment in the present examples, but do not establish a general monotonic-convergence guarantee.

For practical truncation, in the rest of our results, unless otherwise specified, we retain the smallest number $M$ of ranked operators satisfying $\sum_{i=1}^{M}h_{i}^2/\sum_{i=1}^{N}h_{i}^2 \geq0.98$, where $N$ is the total number of candidate operators. This criterion selects 282 of 490 terms for BeH$^+$ and 238 of 476 terms for LiH, corresponding to $57.55\%$ and $50.00\%$ of the candidate sets, respectively. At these cutoffs, the absolute SSPT-D and SSPT errors are $0.27$ and $0.36$~mHa for BeH$^+$, and $0.35$ and $0.15$~mHa for LiH. All four errors lie below the chemical-accuracy threshold of $1.60$~mHa. The cumulative weight is nevertheless a screening heuristic: the candidate states are generally nonorthogonal, and the weights omit the energy denominators and perturbative-space couplings. Consequently, capturing $98\%$ of this weight does not guarantee recovery of $98\%$ of the second-order correction or a specified energy accuracy.

The computational benefit of truncation arises primarily from reducing the number of operator-generated excitation vectors entering the overlap and Hamiltonian matrices. For a transformed Hamiltonian containing $L$ Pauli terms, direct construction of the selected overlap matrix, reference-to-candidate coupling vector, and Hamiltonian matrix involves at most $O(M^2)$, $O(ML)$, and $O(M^2L)$ Pauli expectation-value contributions, respectively. Coefficient-based selection requires sorting the $N$ known weights, with a classical cost of $O(N\log N)$. Coupling-based selection additionally requires the $N$ reference-to-candidate couplings before sorting. Directly expanding these couplings involves at most $O(NL)$ such contributions, unless they have already been computed or cached. Once the $M\times M$ overlap matrix has been constructed, dense canonical orthonormalization and transformation of the selected kernels require $O(M^3)$ classical work in the worst case. The subsequent dense solution of the resolvent equation in the retained rank- $r$ space costs $O(r^3)$, with $r\leq M$. These estimates count algebraic contributions rather than measurement shots and do not imply that each contribution requires an independent measurement. Repeated Pauli products, factorization, caching, and measurement grouping can reduce the measurement requirements.

Overall, coefficient- and coupling-based truncation yields chemically accurate energies with a reduced subset of the candidate operators in these benchmarks, while the SSPT-D results remain sensitive to the diagonal approximation and the changing orthonormal representation. The resolvent treatment gives a more stable correction by retaining perturbative-space couplings, rather than by explicitly regularizing the denominators. Genuine intruder-state problems can therefore persist if $E_{\mathrm{CS}}I-\bar H$ becomes nearly singular, in which case explicit regularization or an enlarged quasi-degenerate model space may be required.

\begin{table*}[t]
\centering
\caption{Summary of several molecules' subspace-SSPT benchmarks using a cumulative 98\% squared-coefficient threshold.}
\label{tab:subspace_sspt_summary}
\begin{threeparttable}
\scriptsize
\setlength{\tabcolsep}{6.0pt}
\begin{tabular}{llccccc}
\toprule
Molecule & Method & $N_{\mathrm{tap}}$ & $q_{\mathrm{chem}}$ & Best error (mHa) & $R_{\mathrm{chem}}$ & $r_{\mathrm{SVD}}$ \\
\midrule
 & CS-VQE & 8 & 5 & 0.186 & 1.00 & -- \\
OH$^-$ & SSPT-D & 8 & 2 & 0.005 & 75.84 & 28 \\
 & SSPT & 8 & 2 & 0.012 & 130.59 & 28 \\
\midrule
 & CS-VQE & 8 & 5 & 0.244 & 1.00 & -- \\
BeH$^+$ & SSPT-D & 8 & 1 & 0.019 & 24.08 & 63 \\
 & SSPT & 8 & 1 & 0.009 & 198.49 & 63 \\
\midrule
 & CS-VQE & 8 & 3 & 0.015 & 1.00 & -- \\
HF & SSPT-D & 8 & 1 & 0.001 & 13.72 & 24 \\
 & SSPT & 8 & 1 & 0.001 & 8.92 & 24 \\
\midrule
 & CS-VQE & 8 & 4 & 0.120 & 1.00 & -- \\
LiH & SSPT-D & 8 & 1 & 0.022 & 16.90 & 58 \\
 & SSPT & 8 & 1 & 0.006 & 73.67 & 58 \\
\midrule
 & CS-VQE & 9 & 6 & 0.684 & 1.00 & -- \\
BeH$_2$ & SSPT-D & 9 & 1 & 0.140 & 97.74 & 66 \\
 & SSPT & 9 & 1 & 0.056 & 490.50 & 66 \\
\midrule
 & CS-VQE & 15 & 6 & 0.152 & 1.00 & -- \\
F$_2$ & SSPT-D & 15 & 1 & 0.030 & 16.58 & 124 \\
 & SSPT & 15 & 3 & 0.034 & 4.18 & 157 \\
\midrule
 & CS-VQE & 16 & 3 & 0.026 & 1.00 & -- \\
HCl & SSPT-D & 16 & 1 & 0.010 & 3.73 & 52 \\
 & SSPT & 16 & 2 & 0.010 & 3.36 & 48 \\
\bottomrule
\end{tabular}
\end{threeparttable}
\medskip
\noindent
\begin{minipage}{\textwidth}
\footnotesize
\raggedright
$N_{\mathrm{tap}}$ is the qubit number after using the $Z_{2}$ tapering. For each contextual-qubit point, the nonzero Pauli terms in $V_{\mathrm{anti}}$ are sorted by descending $h_{i}^2$, and the smallest prefix satisfying $\sum_{i=1}^{M_{\mathrm{PT}}}h_{i}^2/\sum_{i=1}^{N}h_{i}^2\geq 0.98$ is retained. 
$q_{\mathrm{chem}}$ is the minimum contextual-qubit number among the scanned points $q=1,\ldots,7$ satisfying $|\Delta E| \leq 1.6$ mHa. Best error is the smallest absolute error over this scan.
$R_{\mathrm{chem}}$ denotes the ratio of the absolute CS-VQE error to that of the corresponding method (SSPT-D or SSPT), evaluated at the minimum qubit count $q_{\mathrm{chem}}$ required by that method to achieve chemical precision. Larger values indicate greater relative error reduction compared with CS-VQE at the same qubit count.
$N_{\mathrm{exc}}$ is the total number of eligible Pauli excitation terms in $V_{\mathrm{anti}}$ after merging duplicate strings and before the 98\% selection. $M_{\mathrm{PT}}$ and $r_{\mathrm{SVD}}$ are the selected Pauli-term count and retained SVD rank. All three quantities are evaluated at the listed method's $q_{\mathrm{chem}}$ and are shown as ``--'' for CS-VQE.
\end{minipage}
\end{table*}

\subsection{Truncation and benchmarking}
\label{subsec:truncation}

\begin{figure}
    \centering
    \hspace{-5mm}
    \includegraphics[width=0.93\linewidth]{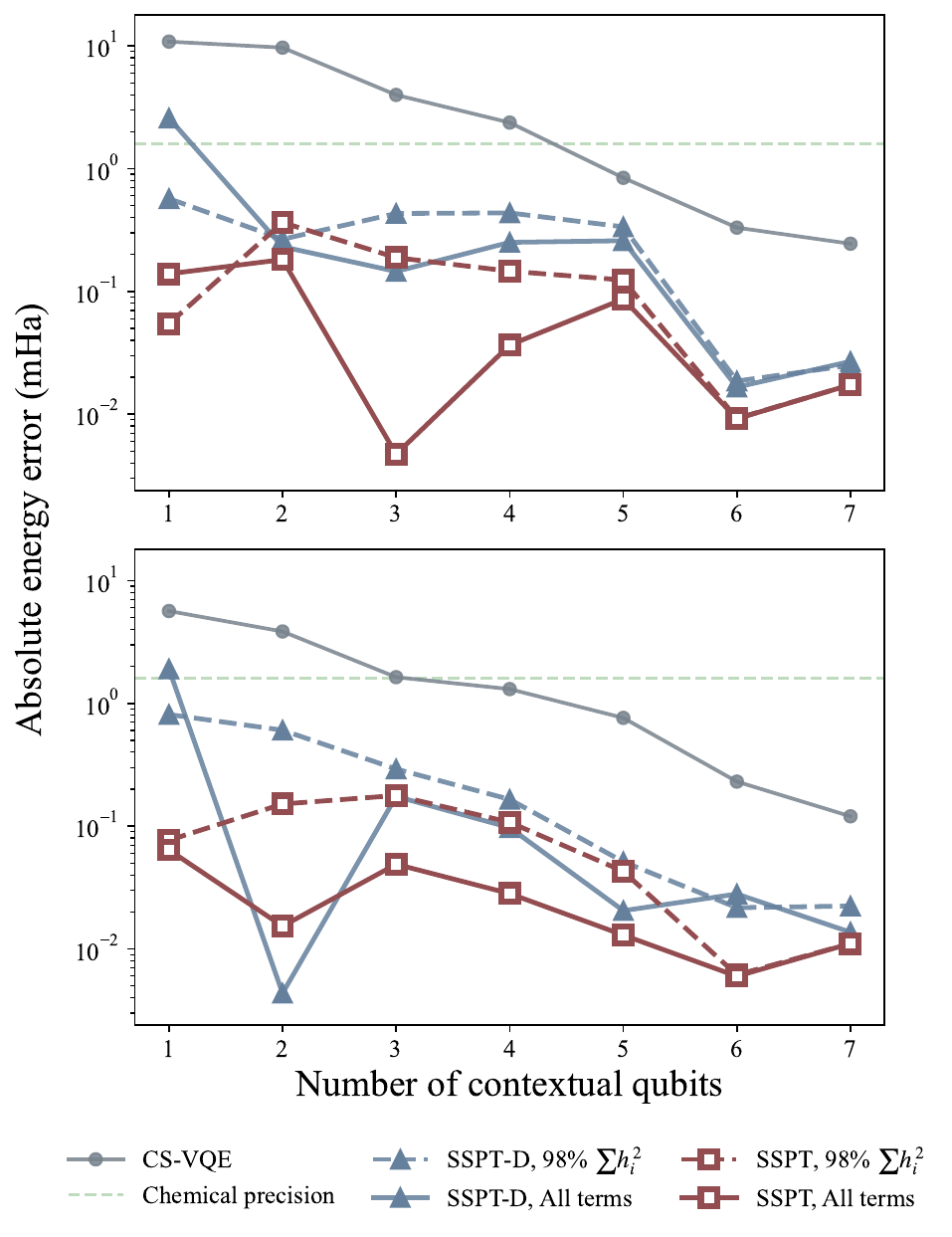}
    \caption{Absolute energy errors of CS-VQE and perturbatively corrected subspace energies for BeH$^+$ (upper) and LiH (lower) with increasing contextual-subspace size. SSPT-D and SSPT use either all Pauli excitation operators or the smallest leading subset accounting for at least 98\% of the cumulative squared coefficient weight in $\widetilde{V}$, ranked by decreasing $h_{i}^2$. }
    \label{fig:subspace_range_qubit}
\end{figure}

The factorized implementation recovers a substantial fraction of the correlation effects lost under contextual projection while avoiding explicit construction of the full-system statevector. Matrix elements involving the excluded stabilizer sectors are evaluated through expectation values on the contextual register and analytically determined contributions from the fixed stabilizer qubits. The resulting perturbative correction therefore retains the reduced quantum-register requirements of CS-VQE while accounting for couplings to configurations outside the retained stabilizer sector. 
To assess the effectiveness of the truncation strategy across different contextual qubit number, we perform calculations for BeH$^+$ and LiH using one to seven contextual qubits, as shown in Fig.~\ref{fig:subspace_range_qubit}. We compare the results obtained by retaining Pauli terms accounting for 98\% of the cumulative squared coefficient weight, $\sum_i |h_i|^2$, with those obtained using the complete set of candidate excitation terms. 
It is worth noting that the results presented here differ from those in Fig.~\ref{fig:range_qubit} in the ordering of the excitation terms, even when all candidate terms are retained. In Fig.~\ref{fig:range_qubit}, the excitation terms follow their original order in the rotated Hamiltonian, without additional sorting, whereas here they are ranked in descending order of their squared Pauli coefficients, $h_{i}^2$.
In general, retaining all candidate terms yields lower energy errors than the 98\% truncation strategy, although a few exceptions are observed. Furthermore, SSPT generally outperforms SSPT-D, suggesting that accounting for off-diagonal couplings within the perturbative subspace provides a more accurate description of the neglected contributions. Nevertheless, the 98\% truncation strategy retains much of the accuracy improvement achieved by the complete candidate set, indicating that a relatively small fraction of $h_{i}^2$ can be discarded without substantially compromising accuracy in most cases. As the number of contextual qubits increases, the differences between the truncated and untruncated results generally diminish, reflecting the reduced significance of the neglected perturbative contributions.

For the extended molecular benchmarks summarized in Table~\ref{tab:subspace_sspt_summary}, candidate perturbative Pauli operators are ranked by the magnitudes of their Hamiltonian coefficients $h_{i}^2$, with a subset of the leading 98\% terms and all of them for comparison. 
Table~\ref{tab:subspace_sspt_summary} reports the tapered qubit count, the minimum contextual-register size required to reach chemical precision for each method, the best absolute error, the improvement factor at the minimum-qubit chemical-accuracy point and the rank obtained from SVD. 
Within the tested range of one to seven contextual qubits, both SSPT-D and SSPT reduce the minimum number of qubits required to achieve chemical precision ($|\Delta E|\leq 1.60$~mHa) for all seven molecules. Specifically, both methods reduce the required qubit count from five to two for OH$^-$, five to one for BeH$^+$, three to one for HF, four to one for LiH, and six to one for BeH$_2$. For F$_2$, SSPT-D and SSPT reduce the requirement from six qubits to one and three, respectively, while for HCl, the corresponding reductions are from three to one and two. Overall, compared with bare CS-VQE, SSPT-D saves two to five contextual qubits, whereas SSPT saves one to five.

Both corrections also reduce the best absolute errors, with the relative improvement over CS-VQE quantified by the error-reduction factor $R_{\mathrm{chem}}$ across the scan. $R_{\mathrm{chem}}=|\Delta E_{\mathrm{CS\text{-}VQE}}(q_{\mathrm{chem}}^{\mathrm{method}})|/|\Delta E_{\mathrm{method}}(q_{\mathrm{chem}}^{\mathrm{method}})|$ measures the relative error reduction at the listed method's chemical-precision point, where $\mathrm{method}$ denotes SSPT-D or SSPT. Here, $|\Delta E_X(q)| = |E_X(q)-E_{\mathrm{FCI}}|$ is the absolute energy error relative to FCI. $R_{\mathrm{chem}}>1$ indicates reduced absolute error and larger ratios mean stronger relative improvement at this point. For BeH$^+$, the best error decreases from $0.24$~mHa for CS-VQE to $0.02$~mHa for SSPT-D and $0.01$~mHa for SSPT, while $R_{\mathrm{chem}}$ is $24.08$ and $198.49$, respectively.
Another particularly large improvements occur for OH$^-$: at the two-qubit chemical-accuracy point, SSPT-D and SSPT reduce the CS-VQE error by factors of $75.84$ and $130.59$, respectively, and obtaining the best error from $0.186$ to $0.005$ and $0.012$~mHa. The smallest errors are obtained for HF, where the best SSPT-D and SSPT errors are $0.75\times10^{-3}$ and $1.38\times10^{-3}$~mHa, respectively.

The resolvent formulation does not uniformly outperform the diagonal approximation under the finite operator budget used here. For OH$^-$ and F$_2$, SSPT-D gives a smaller best error than SSPT. Also, SSPT may require more qubits than SSPT-D to achieve chemical precision for certain molecules, such as F$_2$ and HCl. Nevertheless, at the qubit counts where SSPT-D achieves chemical precision, SSPT still yields relatively small absolute errors of $1.65$~mHa and $2.01$~mHa for F$_2$ and HCl, respectively. These results demonstrate that a limited set of stabilizer-incompatible Pauli operators can provide useful corrections and reduce the number of qubits needed for chemical precision. They do not imply pointwise or monotonic improvement throughout the scan, and the relative performance of SSPT-D and SSPT remains dependent on the molecule, contextual subspace, and selected perturbative operators (even their ordering for SSPT-D).

\subsection{Potential energy curves}
\label{subsec:pec}

\begin{figure*}
    \centering
    \includegraphics[width=0.80\linewidth]{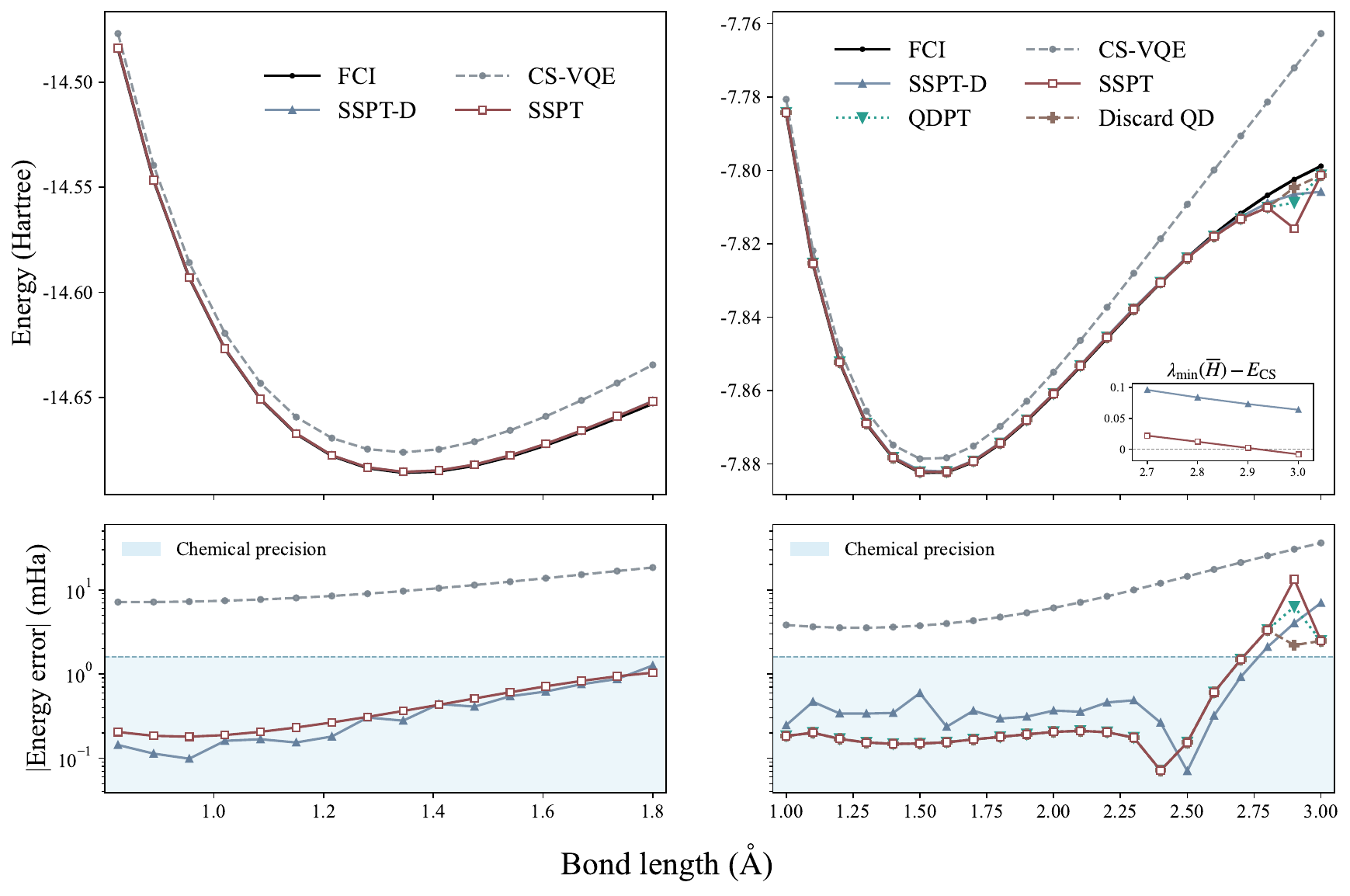}
    \caption{Potential energy curves of BeH$^+$ (left) and LiH (right) using two qubits, comparing bare CS-VQE with perturbative corrections. Upper panels show absolute energies and lower panels show absolute errors relative to FCI on a logarithmic scale.
    The inset shows the signed energy difference $\lambda_{\min}(\overline{H})-E_{\mathrm{CS}}$ in Hartree as a function of the LiH bond length. Here, $E_{\mathrm{CS}}$ is the CS-VQE reference-state energy at the same bond length. SSPT uses the lowest eigenvalue of $\overline{H}$, and SSPT-D uses the smallest diagonal element of $\overline{H}$ in the same orthonormal basis.}
    \label{fig:PEC}
\end{figure*}

To further assess the performance of SSPT, we applied perturbative corrections to the potential energy curves (PECs) of LiH and BeH$^+$.
From Fig.~\ref{fig:PEC} we can see that SSPT gives a more stable error profile across the PECs than the SSPT-D correction. Using the SSPT correction, couplings among the Pauli-generated external configurations are incorporated, leading to smoother and more consistent corrections near the equilibrium region. By contrast, the SSPT-D correction is more sensitive to the geometry-dependent diagonal energy denominators and therefore exhibits larger fluctuations as the electronic structure changes along bond stretching.

However, a notable exception occurs in the dissociation region of the LiH molecule. At an internuclear distance of 2.9 Å, the resolvent method exhibits a pronounced error spike, with the energy error reaches approximately $13.37 ~\text{mHa}$, compared to errors of approximately $3.34~\text{mHa}$  and $2.47~\text{mHa}$ at the adjacent 2.8 Å and 3.0 Å points, respectively. Interestingly, SSPT-D maintains a smaller absolute error (about $4.0~\text{mHa}$) at this specific geometry.

Analysis of the perturbation subspace eigenvalues $\overline{H}$ reveals that this error spike is a manifestation of the classic intruder-state-like problem. At 2.9 Å, as static correlation increases due to bond stretching, an excited state within the perturbative subspace becomes quasi-degenerate with the zeroth-order ground state energy $E_\mathrm{CS}$, exhibiting an energy gap of merely $2.38~\text{mHa}$ (see the inset in the upper-right panel of Fig.~\ref{fig:PEC}). Consequently, the matrix $E_{\mathrm{CS}}I-\overline{H}$ becomes nearly singular, producing an anomalously large resolvent correction. The better behavior of SSPT-D at this geometry should therefore be interpreted as accidental rather than systematically more reliable. By neglecting off-diagonal couplings within the external subspace, SSPT-D does not reproduce the small spectral denominator responsible for the resolvent spike at this geometry.
To diagnose and mitigate this breakdown, we considered two strategies. The first is selective state truncation, in which the quasi-degenerate external state is removed from the perturbative manifold (``Discard QD'' in Fig.(\ref{fig:PEC})). This eliminates the singular contribution and smooths the curve, but it also discards a physically important correlation channel. The second is a quasi-degenerate perturbation theory (QDPT)-inspired model-space enlargement strategy, in which external states satisfying a chosen near-degeneracy criterion are promoted into an enlarged model space and treated nonperturbatively through diagonalization together with the CS-VQE reference, following the general strategy of extended model-space perturbation theories \cite{devarajan2008generalized}. 
The remaining well-separated external states are then treated perturbatively. This treatment is more physically motivated, since it moves the strongly coupled near-degenerate component from the perturbative correction into the model-space sector.
Along the bond-stretching coordinate, we adopt a quasi-degeneracy threshold of 5 mHa, identifying a single quasi-degenerate state at a bond length of $2.9$~\AA. Both the Discard QD and QDPT approaches mitigate the resulting error spike. Notably, when the number of contextual qubits is increased from two to three or four, the SSPT error spike at $2.9$~\AA disappears. The corresponding results for the three- and four-qubit contextual subspaces, together with further details on the implementation and settings of Discard QD and QDPT, are provided in Appendix~\ref{appendix:PEC}.

\subsection{Energy barrier of the decomposition of formaldehyde}
\label{subsec:energy_barrier}

To evaluate the performance of factorized perturbative corrections for chemical reaction barriers, we consider the decomposition of formaldehyde as a representative example and examine the electronic energy difference between the reactant S3 and the transition state TS1 along the primary reaction pathway. Specifically, for each geometry, the active-space Hamiltonian is constructed on the cc-pVDZ basis using the converged CASSCF(8e,8o) orbitals and subsequently mapped onto a 16-qubit representation via the Jordan–Wigner transformation. We retain the smallest leading set reaching a cumulative squared-coefficient weight of 98\%. Unlike the other molecular calculations employing qubit-ADAPT-VQE, the CS-VQE ground states in this case are obtained by classically solving the contextual-subspace Hamiltonians and are subsequently used as reference states.

\begin{figure}[!h]
    \centering
    \hspace{-5mm}
    \includegraphics[width=0.93\linewidth]{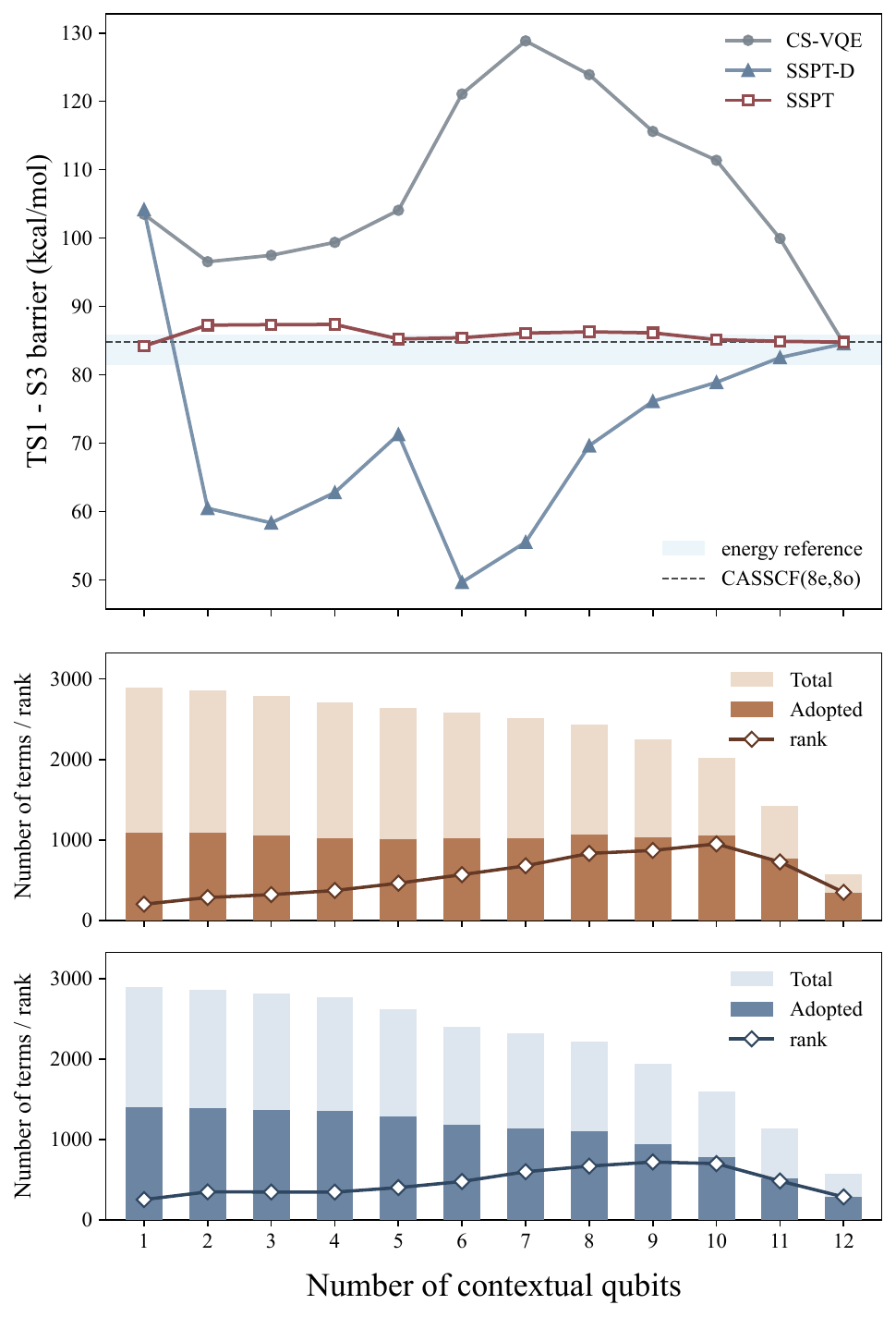}
    \caption{Energy barrier and excitation-space dimensions for H$_2$CO decomposition as a function of the number of contextual qubits. SSPT-D and SSPT employ a cumulative squared-coefficient selection threshold of 98\%. The top panel shows the energy barrier calculated using CS-VQE, SSPT-D, and SSPT. The black dashed line denotes the CASSCF(8e,8o) barrier, while the shaded band (``energy reference'', 81.40--85.94~kcal/mol) represents the range of literature values compiled in~\cite{tao2026measurement}. The middle and bottom panels correspond to S3 and TS1, respectively. Outlined bars show the total number of available excitation terms, and filled bars show the number retained by the 98\% selection. Lines with markers indicate the numerical rank of the matrix formed by the selected excitation states, determined by SVD, which gives the dimension of the linearly independent perturbation space.}
    \label{fig:energy_barrier}
\end{figure}

Taking the CASSCF reaction barrier of 84.79 kcal/mol as the reference, Fig.~\ref{fig:energy_barrier} reveals a particularly pronounced advantage of SSPT over SSPT-D in both accuracy and robustness. SSPT remains close to the reference across the range of qubits, whereas SSPT-D exhibits substantially larger deviations and a stronger dependence on the number of contextual qubits. Across the entire scan, the maximum absolute barrier error (compared to CASSCF) is 2.59 kcal/mol for SSPT, compared with 35.11 kcal/mol for SSPT-D. This difference is particularly evident when using five contextual qubits. The uncorrected CS-VQE barrier is 104.07 kcal/mol, corresponding to an absolute error of 19.28 kcal/mol. Using the 98\% cumulative squared-coefficient threshold, SSPT reduces the barrier to 85.25 kcal/mol, with an absolute error of only 0.47 kcal/mol, whereas SSPT-D yields 71.28 kcal/mol and retains an error of 13.51 kcal/mol. For smaller contextual subspaces, SSPT-D even yields larger absolute barrier errors than bare CS-VQE at some points, e.g., using three contextual qubits, SSPT-D yields larger absolute barrier errors than uncorrected CS-VQE with 26.43 kcal/mol compared to 12.70.

The lower panels in Fig.~\ref{fig:energy_barrier} highlight the distinction between the number of retained perturbation terms and the dimension of the resulting excitation space. Even with a cumulative squared-coefficient threshold of 98\%, only approximately 38–60\% of the available nonzero perturbation Pauli terms are retained, substantially reducing the operator count while preserving most of the coefficient weight. Moreover, similar retained-term counts can correspond to markedly different ranks. For S3, the number of retained terms remains approximately constant over 1–10 qubits, whereas the SVD rank increases from 201 to 950. Specifically, the 1- and 10-qubit calculations retain 1087 and 1058 terms, respectively, yet the rank at 10 qubits is approximately 4.7 times that at 1 qubit. This reflects differences in the linear dependence among the excitation vectors generated by the selected operators acting on the respective reference states. Consequently, the number of retained operators alone does not determine the number of linearly independent excitation directions. The numerical rank provides the relevant measure of the effective perturbative-space dimension.

Increasing the number of perturbative terms does not necessarily lead to a monotonic reduction in the error. This behavior indicates an imbalance in the recovery of correlation energies between the reactant and transition-state geometries. It should also be emphasized that no prior qubit tapering based on $Z_2$ symmetries is applied in these calculations. Consequently, the qubit scan still contains degrees of freedom that could, in principle, be fixed exactly by symmetry. The reaction barrier is already close to the reference value at 12 qubits. For the 13–15-qubit cases, the remaining fixed stabilizers commute with the full Hamiltonian, and the selected symmetry sector is sufficient to recover the target ground state. Consequently, the perturbative coupling between the contextual subspace and its complementary space vanishes. As a result, the CS-VQE, SSPT-D, and SSPT barriers obtained with 13–16 qubits coincide with the CASSCF reference within numerical precision, giving rise to an essentially flat plateau beyond 12 qubits. 

These results demonstrate that, for the present system, an appropriate perturbative correction can accurately recover the reaction barrier within the active space while requiring fewer explicitly simulated qubits than the full 16-qubit representation. 

\section{Discussion}
In this work, we developed Stabilizer-Subspace Perturbation Theory (SSPT) as a perturbative framework for recovering correlation effects neglected by stabilizer-based subspace projections, with CS-VQE serving as a representative application. The central idea is to exploit the Pauli terms eliminated by the stabilizer projection to construct an external perturbative space, rather than treating their contributions as an unrecoverable projection error. By applying these operators to the optimized CS-VQE reference state, SSPT incorporates the effects of couplings between the retained contextual subspace and its orthogonal complement through second-order perturbative corrections. Furthermore, the factorized structure of the stabilizer-constrained reference state allows the required matrix elements to be evaluated using Pauli expectation values within the reduced contextual register, without explicitly reconstructing the full Hilbert space.

Numerical benchmarks demonstrate that SSPT can substantially improve the accuracy of CS-VQE while preserving its qubit-saving advantage. For LiH and BeH$^+$, both perturbative formulations significantly reduce the energy errors and, in several cases, achieve chemical precision with fewer contextual qubits than bare CS-VQE. Similar improvements are observed across a broader set of molecular systems. Calculations of potential energy curves and reaction barriers further demonstrate the applicability of SSPT to chemically relevant energy differences. In particular, the perturbative corrections substantially improve the predicted formaldehyde decomposition barrier relative to the reference calculated within the same active space. The method also reduces the size-consistency errors observed in the separated H$_2$--H$_2$ benchmark (see Appendix~\ref{appendix:size_consistency}). Together, these results indicate that the Pauli terms discarded during stabilizer projection contain physically relevant correlation contributions that can be recovered without enlarging the variational quantum register.

The two perturbative formulations exhibit distinct numerical characteristics. The SSPT-D correction provides a computationally simple approximation but can be sensitive to small energy denominators and may overestimate the recovered correlation energy. In contrast, the SSPT retains couplings within the selected external space and generally provides more stable results. Nevertheless, neither approach is immune to near-degeneracy effects. The error spike observed in the stretched LiH potential energy curve illustrates an intruder-state-like problem arising when an external state approaches the CS-VQE reference energy. The reliability of SSPT therefore depends on both the quality of the variational reference state and the selection and conditioning of the perturbative space. Improved operator selection and treatments of near-degenerate configurations remain important directions for further development.

More broadly, SSPT demonstrates how the algebraic structure of stabilizer projections can be exploited to construct physically motivated perturbative spaces, providing a connection between quantum resource reduction and correlation recovery. This strategy may be extended to other stabilizer-constrained variational methods and is naturally compatible with hybrid quantum-classical computing architectures, where quantum processors prepare correlated reference states and classical computations assist in evaluating perturbative corrections. Recent advances in quantum-centric supercomputing further highlight the potential of such hybrid approaches for challenging electronic structure problems~\cite{merz2026crossing,robledo-moreno2025chemistry}. Future developments in perturbative-space selection, measurement efficiency, and near-degeneracy treatments may extend the applicability of SSPT to larger molecular systems, offering a promising direction for resource-efficient quantum chemistry simulations.

\begin{acknowledgments}

This work is supported in part by the National Natural Science Foundation of China under Grant Nos. 62304037, 22422304, and 22393913, in part by the Jiangsu Provincial Scientific Research Center of Applied Mathematics under Grant No. BK20233002, in part by the Start-up Research Fund of Southeast University under Grant No. RF1028623173, in part by the robotic AI-Scientist platform of the Chinese Academy of Sciences, and in part by the Postgraduate Research and Practice Innovation Program of Jiangsu Province under Grant No. 26CXJH0642.

\end{acknowledgments}

\section*{Data Availability}
The data and code supporting the findings of this study are openly available~\cite{yao2026sspt}.

\appendix
\section{Size consistency}
\label{appendix:size_consistency}

\begin{figure}[!h]
    \centering
    \includegraphics[width=0.8\linewidth]{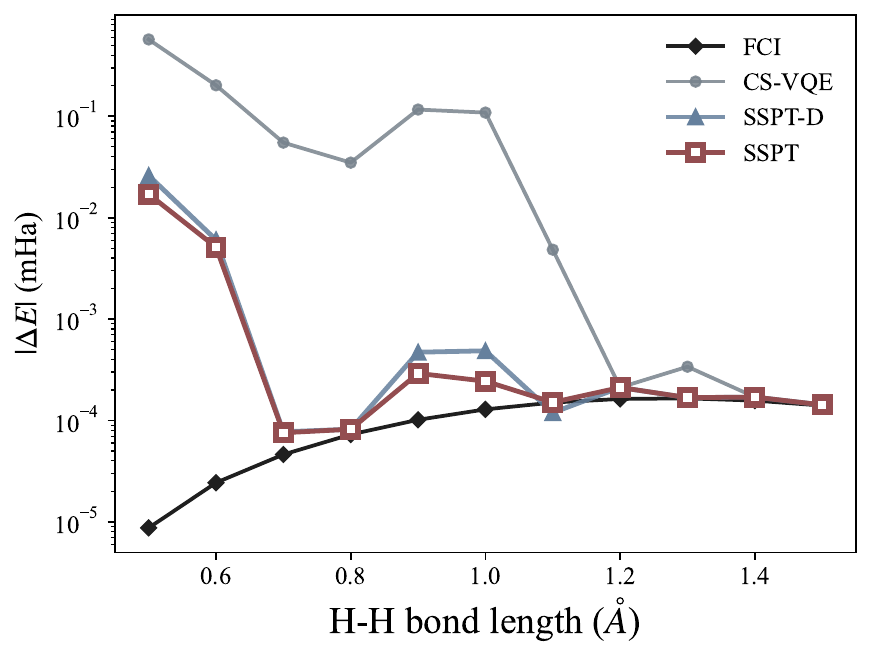}
    \caption{Size-consistency error for two H$_2$ fragments separated by 12 Angstrom.}
    \label{fig:size_consistency}
\end{figure}

We further assessed the size-consistency of the SSPT subspace scheme using a separated $\mathrm{H}_{2}-\mathrm{H}_{2}$ benchmark at $R_{\text{sep}}=12~\text{\AA}$ (see Fig.(\ref{fig:size_consistency})). The size-consistency error is defined as $\Delta E=E(\mathrm{H}_{2}+\mathrm{H}_{2})-2E(\mathrm{H}_{2})$, for which should vanish for an exactly size-consistent method in the noninteracting-fragment limit.

Throughout the full bond-length scan ($r=0.5$ to $1.5~\text{\AA}$), the FCI reference remains essentially additive, with residuals only on the order of $10^{-5}-10^{-4}$ mHa, confirming that the fragments are effectively noninteracting at this separation.
In contrast, bare CS-VQE shows a larger, although still sub-mHa, size-consistency defect. The maximum error reaches 0.57 mHa at the most compressed geometry, and the CS-VQE curve remains systematically above the FCI baseline throughout the scan.
By comparison, both perturbative corrections substantially reduce this non-additivity. The SSPT-D and SSPT energy errors remain below $10^{-1}$ mHa for all tested geometries and, over most of the physically relevant bond-length range, decrease to the $10^{-4}$ mHa scale, close to the FCI benchmark. The SSPT correction is marginally more robust than SSPT-D in the short-bond regime, where the size-consistency defect is most pronounced.
These results indicate that the subspace SSPT-D and SSPT largely preserves the separability of distant fragments and, therefore, yields a practically size-consistent post-CS-VQE description.

\section{Basis freezing threshold}
\label{appendix:basis_freezing_threshold}

\begin{figure*}
    \centering
    \includegraphics[width=0.9\linewidth]{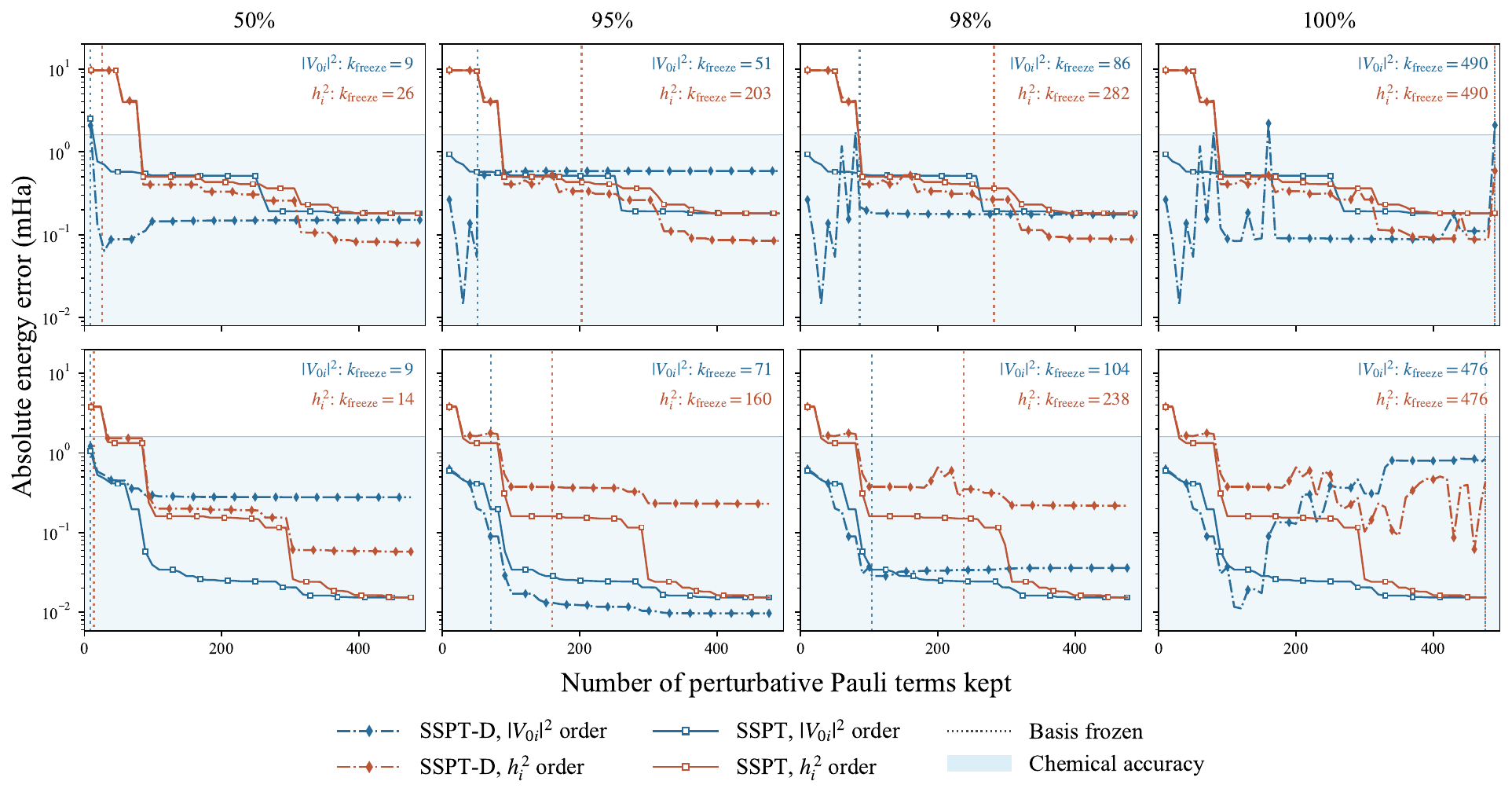}
    \caption{Comparison of SSPT and SSPT-D energy errors under different basis-freezing thresholds. The upper and lower rows show the results for BeH$^+$ and LiH, respectively. The percentages above each column indicate the corresponding freezing thresholds (50\%, 95\%, 98\%, and 100\%).
}
    \label{fig:freezing_threshold}
\end{figure*}

We examine freezing thresholds of 50\%, 95\%, 98\%, and 100\% to assess their influence on convergence under perturbative-term truncation. SSPT retains the off-diagonal elements of the projected Hamiltonian and is therefore invariant under basis rotations within a fixed excitation subspace. Consequently, changing the freezing threshold does not affect the SSPT correction provided that the retained subspace remains unchanged, consistent with the identical full-term limits observed here. The diagonal approximation in SSPT-D introduces a dependence on the basis representation. For both BeH $^+$ and LiH, ordering by $h_i^2$  yields smaller variations in the final SSPT-D energy across the thresholds examined than ordering by the coupling strength $|V_{0 i}|^2$. For BeH $^+$, the $h_i^2$ ordering also gives smaller final energy errors at all four thresholds when all perturbative terms are included. For LiH, however, it produces larger residual errors than the coupling-based ordering at the 95\% and 98\% thresholds, although these errors remain well within chemical precision.

The selected thresholds correspond to substantially different numbers of Pauli terms, allowing convergence to be examined across different degrees of truncation. For the two molecules and both orderings considered here, the 50\%, 95\%, and 98\% thresholds select approximately 2–5\%, 10–41\%, and 18–58\% of the available terms, respectively, whereas the 100\% case includes the complete term set. These choices therefore cover a broad range of retained-term counts. In particular, the 98\% criterion retains at least 98\% of the cumulative squared weight while requiring much fewer terms than the complete set. Compared with the 95\% criterion, it also includes additional contributions with smaller ranking weights. We therefore adopt 98\% as the default truncation criterion in most of this work, as a practical compromise between cumulative-weight retention and the number of Pauli terms required.

\section{Analysis of LiH's potential energy surface}
\label{appendix:PEC}

\begin{figure}[!h]
    \centering
    \hspace{-5mm}
    \includegraphics[width=0.95\linewidth]{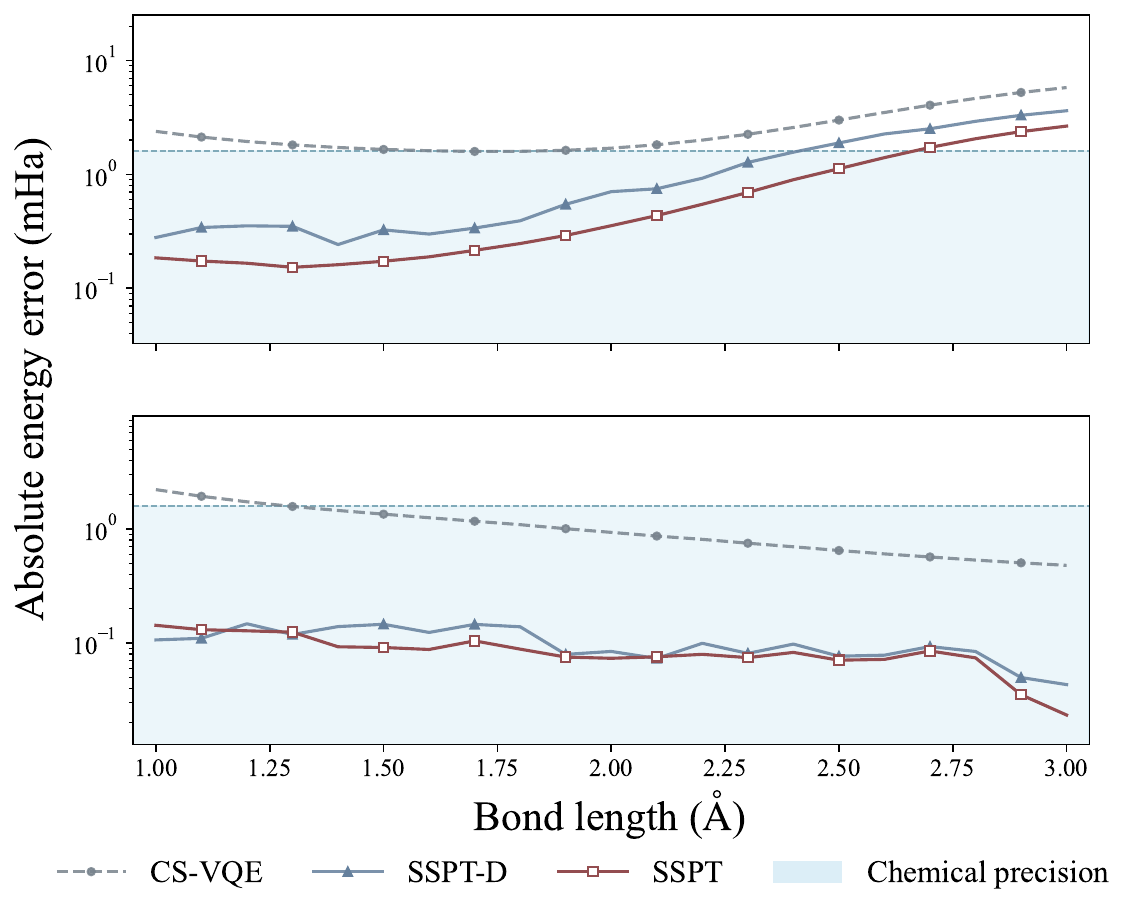}
    \caption{Absolute energy errors relative to the
    FCI reference along the LiH bond dissociation
    coordinate, using three (upper panel) and four
    (lower panel) contextual qubits.}
    \label{fig:pec34}
\end{figure}

In the dissociation regime, the SSPT correction may become sensitive to quasi-degenerate states whose energies approach that of the CS-VQE reference state, leading to small energy denominators and potentially anomalous energy corrections. To investigate and mitigate this behavior, we consider two alternative treatments, denoted as Discard QD and QDPT (quasi-degenerate perturbation theory). Both approaches employ the same perturbative subspace but differ in how the quasi-degenerate states are incorporated into the energy correction.

Recall Eq.~(\ref{eq:orth_matrices})
\begin{equation}
\overline{H}
=
X^\dagger H_{\mathrm{pt}}X,
\qquad
\overline{\mathbf{b}}
=
X^\dagger\mathbf{b}.
\end{equation}
We diagonalize the projected Hamiltonian, obtaining its eigenvalues and eigenvectors
\begin{equation}
    \overline{H} u_j=\varepsilon_j u_j.
\end{equation}
The corresponding couplings
to the CS-VQE reference state are
\begin{equation}
    g_j=u_j^\dagger\overline{\mathbf{b}}.
\end{equation}
We identify quasi-degenerate states according to the energy criterion
\begin{equation}
    \mathcal D=\left\{j:\left|E_{\mathrm{CS}}-\varepsilon_j\right|<\delta_{\mathrm{QD}}\right\},
\end{equation}
where $\delta_{\mathrm{QD}}=5$ mHa is used for LiH calculations. This criterion depends solely on the energy separation, irrespective of the coupling strength $g_j$.

Both treatments retain the same second-order contribution from the non-quasi-degenerate external states,
\begin{equation}
    \Delta E_{\mathrm{nonQD}}^{(2)} =\sum_{j\notin\mathcal D}\frac{|g_j|^2}{E_{\mathrm{CS}}-\varepsilon_j}.
\end{equation}

In the Discard QD approach, the contributions from states in $\mathcal D$ are simply excluded from the perturbative correction, yielding
\begin{equation}
    E_{\mathrm{Discard}}=E_{\mathrm{CS}}+\Delta E_{\mathrm{nonQD}}^{(2)}.
\end{equation}
This procedure removes the potentially divergent contributions associated with small denominators but neglects the coupling between the reference state and the discarded quasi-degenerate states.

In contrast, QDPT explicitly incorporates these couplings by constructing a reduced Hamiltonian in the model space spanned by the reference state and the quasi-degenerate external states,
\begin{equation}
    H_{\mathrm{model}}=
\begin{pmatrix}
E_\mathrm{CS} & \mathbf g_{\mathcal D}^{\dagger}\\
\mathbf g_{\mathcal D} &
\operatorname{diag}(\boldsymbol{\varepsilon}_{\mathcal D})
\end{pmatrix},
\end{equation}
where $\mathbf g_{\mathcal D}$ and $\boldsymbol{\varepsilon}_{\mathcal D}$ collect the couplings and energies of the quasi-degenerate states, respectively. The corrected energy is then evaluated as
\begin{equation}
    E_{\mathrm{QDPT}}=\lambda_{\min}(H_{\mathrm{model}})+\Delta E_{\mathrm{nonQD}}^{(2)},
\end{equation}
where $\lambda_{\min}$ is the minimum eigenvalue of $H_{\mathrm{model}}$.
Here, the non-quasi-degenerate second-order correction is evaluated using the original reference energy and couplings, without further self-consistent updates. When no quasi-degenerate states are identified, both approaches reduce to the original SSPT result.

The advantage of explicitly treating quasi-degenerate states can be illustrated by considering a single external state with energy $\varepsilon$ and coupling $g$. Its conventional second-order contribution, $|g|^2/(E_\mathrm{CS}-\varepsilon)$, becomes highly sensitive as $\varepsilon$ approaches $E_\mathrm{CS}$. In contrast, diagonalization of the two-state model yields
\begin{equation}
    \lambda_-=\frac{E_\mathrm{CS}+\varepsilon}{2}-\sqrt{\frac{(E_\mathrm{CS}-\varepsilon)^2}{4}+|g|^2},
\end{equation}
which remains finite in the degenerate limit, where $\lambda_-=E_\mathrm{CS}-|g|$. Thus, while Discard QD suppresses the small-denominator instability by omitting the corresponding contributions, QDPT retains the coupling-induced energy shifts through explicit diagonalization of a reduced model Hamiltonian constructed from the reference and quasi-degenerate external states. This provides an approximate quasi-degenerate extension of SSPT following the diagonalization of the external-space Hamiltonian.

To further assess the performance of SSPT and SSPT-D beyond the two-qubit contextual subspace considered in the main text, we extend the LiH calculations to three and four contextual qubits. As shown in Fig.~\ref{fig:pec34}, increasing the contextual subspace size improves the accuracy of CS-VQE, with the three-qubit calculation reaching chemical precision at several bond lengths near equilibrium and the four-qubit calculation satisfying this threshold over most of the investigated range. Notably, the three-qubit CS-VQE error exhibits a shallow minimum at intermediate bond lengths before increasing again, whereas the four-qubit error decreases approximately monotonically with increasing bond length. This difference indicates that expanding the contextual subspace not only reduces the overall error but also changes its dependence on molecular geometry. Both perturbative approaches consistently reduce the errors relative to CS-VQE, with SSPT generally outperforming SSPT-D. In particular, for four contextual qubits, both methods achieve chemical precision across the entire bond-length range.
Interestingly, the pronounced error spike observed for the two-qubit SSPT calculation is absent in both the three- and four-qubit results. This observation suggests that the near-degenerate coupling responsible for the anomalous behavior is suppressed as the contextual subspace is expanded.

\bibliography{references.bib}

\end{document}